\documentclass[aps,prb,twocolumn,showpacs,nofootinbib]{revtex4-1}
\usepackage{natbib}
\usepackage{graphicx}
\usepackage[caption=false]{subfig}
\usepackage{float}
\usepackage{longtable}
\usepackage{bm}
\usepackage{siunitx}
\usepackage{xfrac}
\usepackage{amssymb,mathtools,mathrsfs,bbold,dsfont}
\usepackage{amstext,braket,amsmath,tensor}  
\usepackage{array} 
\usepackage{siunitx}
\usepackage[normalem]{ulem}
\usepackage{empheq}
\usepackage{xcolor}
\usepackage{multirow}
\usepackage{amsthm,etoolbox}
\usepackage{cleveref}

\newcommand{\bq}{\boldsymbol{q}}

\newcommand{\angs}{\textup{\AA}}

\DeclareMathAlphabet{\mathbbmsl}{U}{bbm}{m}{sl}

\DeclareSymbolFont{bbold}{U}{bbold}{m}{n}
\DeclareSymbolFontAlphabet{\mathbbold}{bbold}

\newcommand{\2}{{}_2}
\newcommand{\CDW}{\scriptscriptstyle{\text{CDW}}}
\newcommand{\Exp}{\scriptscriptstyle{\text{Exp}}}

\newcommand{\mH}{\scriptscriptstyle{\text{1H}}}
\newcommand{\ThDFT}{\scriptscriptstyle{\text{Th-DFT}}}
\newcommand{\Th}{\scriptscriptstyle{\text{Th}}}
\newcommand{\HH}{\scriptscriptstyle{\text{2H}}}

\begin{document}

\title{Quantum enhancement of charge density wave in NbS$_2$ in the 2D limit}
\author{Raffaello Bianco$^{1,2,3}$}
\email{rbianco@caltech.edu}
\author{Ion Errea$^{4,5,6}$}
\author{Lorenzo Monacelli$^{3}$}
\author{Matteo Calandra$^{7}$}
\author{Francesco Mauri$^{2,3}$}

\affiliation{$^{1}$ Department of Applied Physics and Materials Science,
California Institute of Technology, Pasadena, California 91125}
\affiliation{$^{2}$ Graphene Labs, Fondazione Istituto Italiano di Tecnologia, Via Morego, I-16163 Genova, Italy}
\affiliation{$^{3}$ Dipartimento di Fisica, Universit\`a di Roma La Sapienza, 
Piazzale Aldo Moro 5, I-00185 Roma, Italy}
\affiliation{$^4$Fisika Aplikatua 1 Saila, Gipuzkoako Ingeniaritza Eskola, University of the Basque Country (UPV/EHU),
Europa Plaza 1, 20018, Donostia-San Sebasti\'an, Basque Country, Spain}
\affiliation{$^5$Centro de F\'isica de Materiales (CSIC-UPV/EHU), 
            Manuel de Lardizabal pasealekua 5, 20018 Donostia-San Sebasti\'an,
	    Basque Country, Spain}
\affiliation{$^6$Donostia International Physics Center (DIPC), 
            Manuel de Lardizabal pasealekua 4, 20018 Donostia-San Sebasti\'an,
	    Basque Country, Spain}	    
\affiliation{$^7${Sorbonne Universit\'e, CNRS, Institut des
  Nanosciences de Paris, UMR7588, F-75252, Paris, France}}

\begin{abstract}
{At ambient pressure, bulk 2H-NbS$_2$ displays no charge density wave instability  at odds with the isostructural and isoelectronic
compounds 2H-NbSe$_2$, 2H-TaS$_2$ and 2H-TaSe$_2$, and in disagreement with harmonic calculations.
Contradictory experimental results have been reported in supported single layers, as 1H-NbS$_2$ on Au(111)
does not display a charge density wave, while 1H-NbS$_2$ on 6H-SiC(0001) endures a $3\times 3$ reconstruction.
Here, by carrying out  quantum anharmonic calculations from first-principles, we evaluate the temperature dependence of phonon spectra in NbS$_2$ bulk
and single layer as a function of pressure/strain. For bulk 2H-NbS$_2$, we find excellent agreement with inelastic X-ray spectra and demonstrate 
the  removal of charge ordering due to anharmonicity.  In the 2D limit, we find an enhanced tendency toward charge density
wave order. Freestanding 1H-NbS$_2$ undergoes a $3\times3$ reconstruction, in agreement with data on 6H-SiC(0001) supported samples. Moreover, as
strains smaller than $0.5\%$ in the lattice parameter are enough to completely remove the $3\times3$ superstructure,
deposition of 1H-NbS$_2$ on flexible substrates or a small charge transfer via field-effect could lead to devices with dynamical switching on/off of
charge order.}
\end{abstract}

\maketitle

\newtoggle{draft}
\togglefalse{draft}

\begin{figure}[t!]
\centering
\includegraphics[width=\columnwidth]{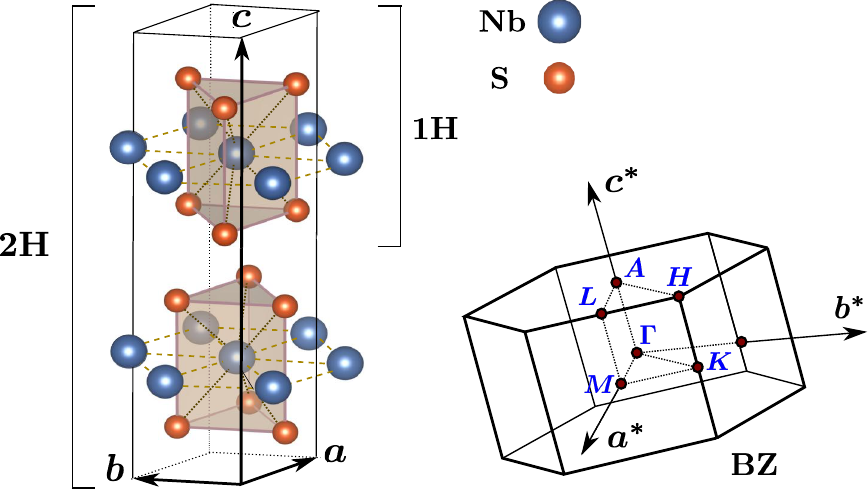}
\caption{(Color online)  Left-hand side: crystal structure of trigonal NbS$\2$ in the 1H 
monolayer and in the 2H (bulk) stacking layer configuration. Right-hand side: 
Corresponding hexagonal Brillouin zone ($BZ$) with the high-symmetry points (in the monolayer 
configuration only the points $\Gamma M K$ are relevant, and they are customarily 
indicated with a line over the letter).}
\label{fig:crystal}
\end{figure}

Transition metal dichalcogenides (TMDs) are layered materials  with generic formula MX${}_2$,  where M is a transition metal (Nb, Ta, Ti, Mo, W, \dots) 
and X a chalcogen (S, Se, Te). The layers, made of triangular lattices of transition metal atoms sandwiched
by covalently bonded chalcogens, are held together by weak van der Waals forces, and TMDs can be readily exfoliated into thin flakes down to the single layer limit, with mechanical or chemical techniques~\cite{Novoselov10451,Mak,Rad,Zhiyuan}. 
In TMDs, the interplay between strong electron-electron and  electron-phonon 
interactions gives rise to rich phase diagrams, with a wide variety of cooperating/competing collective electronic orderings as charge-density wave (CDW), 
Mott insulating, and superconductive phases~\cite{doi:10.1080/00018737500101391,PhysRevLett.121.026401}.
{Of the several polytypes, we focus here on the most common one for NbS$_2$, the H polytype~\cite{PhysRevB.86.155125,PhysRevB.97.195140}, where the transition metal is in trigonal prismatic coordination with the surrounding chalcogens. In Fig.~\ref{fig:crystal} the 1H (monolayer) and 2H (bulk) crystal structures are shown.}

Among metallic 2H bulk TMDs, NbS$_2$ occupies a special place since no CDW has been reported~\cite{FisherSienko,note1},
contrary to its isoelectronic and isostructural 2H-TaSe$\2$, 2H-TaS$\2$ and  2H-NbSe$\2$. All these systems have very similar band structures and are conventional (i.e. phonon-mediated)
superconductors with critical temperatures $T_c$ that increases from a sub-Kelvin value in 2H-TaSe$\2$ and 2H-TaS$\2$ (around $0.2$~K and $0.5$~K, respectively) 
up to $5.7$~K  in 2H-NbS$\2$ and $7.2$~K in 2H-NbSe$\2$~\cite{Moratalla,PhysRevB.78.104520,PhysRevB.15.2943,PhysRevLett.119.087003}. 
They also show quite a different CDW transition strength~\cite{doi:10.1143/JPSJ.51.219,PhysRevLett.86.4382}. 2H-TaSe$\2$, 2H-TaS$\2$ and 2H-NbSe$\2$ undergo a triple incommensurate CDW transition to a superlattice with hexagonal symmetry corresponding roughly to the same wave-vector $\bq_{\CDW}=\Gamma\,M(1-\delta)2/3$ ($\delta\simeq 0.02$ is the \textit{incommensurate factor}) of the Brillouin zone. However, the transition temperature $T_{\CDW}$ increases from $30$~K for 2H-NbSe$\2$ to $80$~K for 2H-TaS$\2$ and $120$~K for 2H-TaSe$\2$ (2H-TaSe$\2$ actually shows a further commensurate first-order CDW transition at $92$~K with $\delta$ dropping continuously to zero)~\cite{PhysRevB.16.801}. Therefore, 2H-NbS$\2$ considerably stands out as it shows only an incipient instability near $\bq_{\CDW}$, but it remains stable even at the lowest temperatures. This circumstance is even more surprising if it is considered that 2H-NbSe$\2$ and 2H-NbS$\2$ display superconductivity at similar temperatures.

In TMDs, the behavior of the CDW ordering in the two-dimensional limit cannot be inferred from the knowledge of their bulk counterparts, since two competing mechanisms are expected to play a major role. On the one hand, reduced dimensionality strengthens Peierls instabilities (due to Fermi surface nesting) and electron-phonon interactions (due to reduced dielectric screening), thus favoring stronger CDW. On the other hand, stronger fluctuation effects from both finite temperatures and disorders should tend to destroy long-range CDW coherence in low-dimensional systems~\cite{Xi2015}.  In particular, the effect of dimensionality on the CDW ordering in the H polytype is a current active research area. In 1H-TaS$_2$, the CDW vanishes in the 2D limit~\cite{PhysRevB.98.035203}, while in 1H-TaSe$_2$ it remains unchanged with respect to the bulk~\cite{doi:10.1021/acs.nanolett.7b03264}. For 1H-NbSe$_2$ and 1H-NbS$_2$ the situation is more debated.
In the 1H-NbSe$_2$ case, $3\times 3 $ CDW is observed, but  some controversy is still present in literature,  tentatively attributed either to the sample exposure to air or to  the different substrates, concerning the thickness dependence of the $T_{\CDW}$ (lower/higher $T_{\CDW}$ of the monolayer with respect to the bulk has been reported with bilayer graphene~\cite{Ugeda2015}/silicon~\cite{Xi2015} substrate, respectively). Supported single layers of 1H-NbS$_2$  have become recently available, and while  no traces of CDW have been observed down to $30$~K for monolayers grown on top of Au(111)~\cite{2019arXiv190103552S}, a $3\times3$  CDW ordering  has been observed at ultra-low temperature (measurements performed below $5$~K) for monolayers grown on top of graphitized 
6H-SiC(0001)~\cite{LinNbS2}.

\begin{figure}[t!]
\centering
\includegraphics[width=\columnwidth]{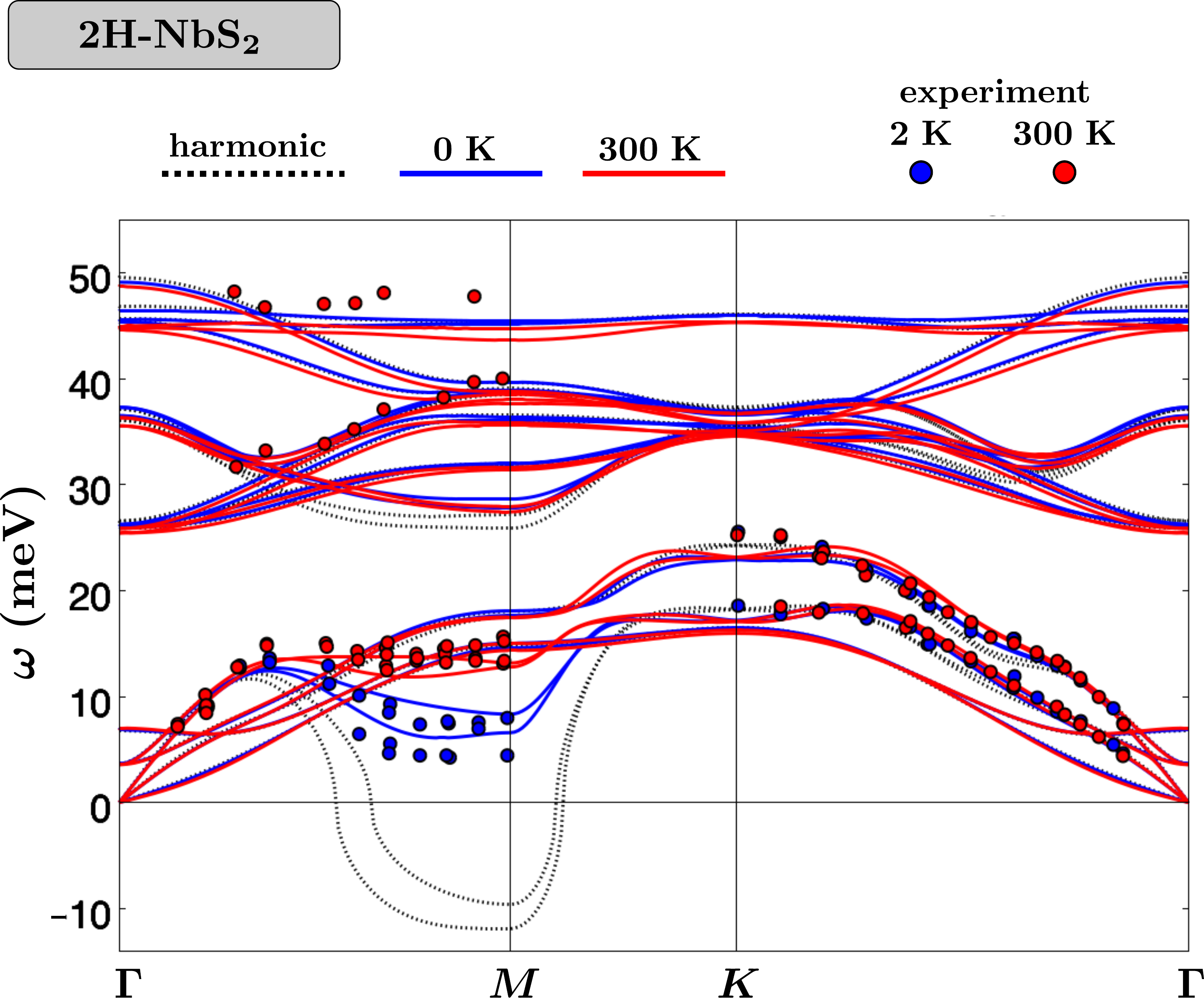}
\caption{(Color online) 2H-NbS$\2$ harmonic (black dashed lines) and SSCHA anharmonic phonon dispersion at $300$~K (red solid lines) and $0$~K (blue solid lines), calculated using the experimental lattice parameters. 
The results are compared with the IXS measures of Ref.~\citenum{PhysRevB.86.155125} performed at $300$~K (red dots) and $2$~K (blue dots). 
The SSCHA dispersion corrects the errors of the pure harmonic result near $M$: the instability of the two longitudinal acoustic and optical modes is removed and  the softening on lowering temperature is well reproduced.}
\label{fig:2H-NbS2_Exp}
\end{figure}

\begin{figure}[t!]
\centering
\includegraphics[height=0.733\textheight, width=\columnwidth]{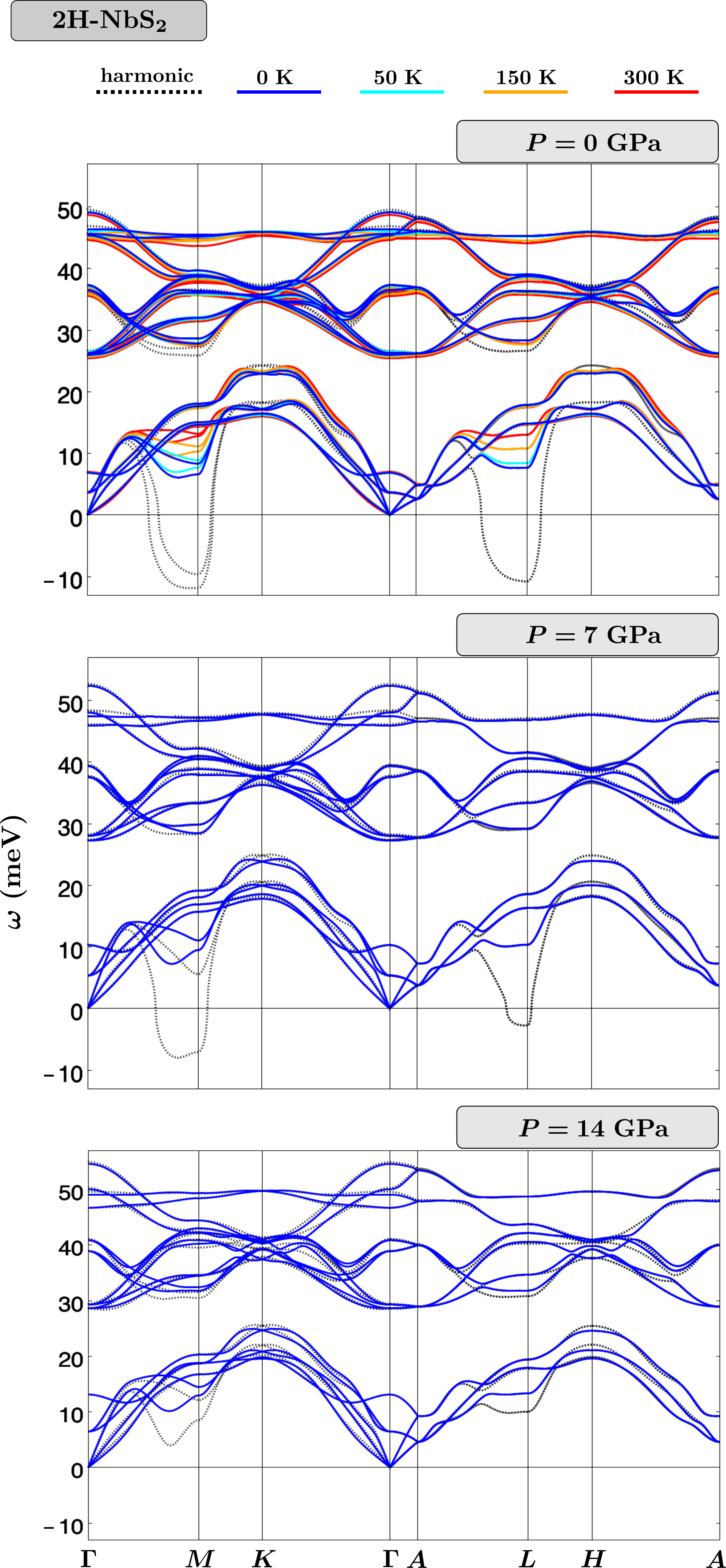} 
\caption{(Color online) 2H-NbS$\2$ harmonic phonon dispersion (black
  dashed lines) and SSCHA anharmonic phonon dispersion at several
  temperatures (colored solid lines). Results for different pressures are
  shown. From the top to the bottom panel: $0$~GPa, $7$~GPa,
  $14$~GPa. The zero pressure results are obtained using the experimental lattice parameters. The high pressure results
  are obtained assuming that the ratio between experimental and DFT theoretical lattice parameters are independent of the 
  applied pressure (more details in the main text). Anharmonicity removes the instability, obtained at harmonic
  level, of the longitudinal acoustic and optical modes near $M$ and
  $L$ at  $0$~GPa and $7$~GPa. Anharmonicity reduces as the pressure
  increases but  it has a noticeable effect even at $14$~GPa.} 
\label{fig:2H-NbS2_Press}
\end{figure}


{In this letter we investigate, from first-principles, the vibrational properties of bulk 2H-NbS$\2$ (at zero and finite pressure) and suspended 1H-NbS$_2$, 
taking into account quantum anharmonic effects at non-perturbative level in the framework of the stochastic self-consistent harmonic approximation (SSCHA)
~\cite{PhysRevB.89.064302,PhysRevB.96.014111,PhysRevB.98.024106,note2}. For bulk 2H-NbS$_2$,  we show
that quantum anharmonic effects remove the instability found at harmonic level, and give 
temperature dependent phonon energies in quantitative agreement with experiment.
Previous anharmonic calculations for 2H-NbS$_2$ anticipated the role of anharmonicity, 
but were limited to a low dimensional subspace of the total high dimensional configurations space and did not account for the
temperature dependence~\cite{PhysRevLett.119.087003}. We also show that quantum anharmonic effects are noticeable even at high pressure. Moreover, we 
demonstrate that the difference between 2H-NbS$\2$ and 2H-NbSe$\2$ is not simply ascribable 
to the different chalcogen mass. 
{Finally, we analyze the 2D limit and show that freestanding single-layer 1H-NbS$_2$ 
undergoes a $3\times3$ CDW instability in agreement with data on 
6H-SiC(0001) supported samples. However, strains smaller than $0.5\%$ 
are sufficient to completely remove the instability, suggesting a strong 
dependence of the CDW on the environmental conditions (substrate, charge transfer...) 
and reconciling the apparent contradiction with supported Au(111) samples.}



For bulk 2H-NbS$\2$, in Fig.~\ref{fig:2H-NbS2_Exp} we compare the computed anharmonic phonon dispersions  
with the results of the inelastic X-ray scattering (IXS) experiment of Ref.~\citenum{PhysRevB.86.155125}, 
at low and ambient temperature. We also show the (temperature-independent) harmonic phonon dispersion. 
Calculations were performed with the $a^{\HH}_{\Exp}=b^{\HH}_{\Exp}=3.33\,\angs$ and $c^{\HH}_{\Exp}=11.95\,\angs$ bulk 
experimental lattice parameters at zero pressure~\cite{PhysRevB.86.155125}.
The phonon dispersion is almost everywhere well reproduced with the harmonic calculation, 
except close to $M$, where it predicts that two longitudinal acoustic and
optical modes {become} imaginary.  Experimental phonon energies show a sensible temperature dependence in this region of the $BZ$ and 
are, obviously, always real. The SSCHA cures the pathology of the harmonic result: 
the anharmonic phonon dispersions do not show any instability and 
give a very good agreement with the experiment at both temperatures.


Since SSCHA calculations give dispersions in good agreement with experiments,
we can perform a wider analysis. In the upper panel of Fig.~\ref{fig:2H-NbS2_Press} 
we show the SSCHA phonon dispersion for different temperatures along 
the full high-symmetry path of the $BZ$. As temperature decreases, anharmonicity causes 
the softening of two acoustic and optical longitudinal modes close to both $M$ and $L$, 
but there is no instability. Thus, quantum fluctuations strongly affected
by the anharmonic potential stabilize 2H-NbS$_2$. In the other two panels we show the effect of 
hydrostatic pressure on the phonon dispersion. Since there are no available experimental lattice parameters at high pressures,
we estimated them by assuming that the ratio between experimental and standard DFT theoretical lattice parameters (i.e. the lattice parameters that minimize the DFT energy but do not take into account
any lattice quantum dynamic effects), $a^{\HH}_{\Exp}(P)/a^{\HH}_{\ThDFT}(P)$ and $c^{\HH}_{\Exp}(P)/c^{\HH}_{\ThDFT}(P)$, are independent of the applied pressure $P$. 
Thus we computed those ratios at zero pressure and, for a given pressure $P$, 
the calculations were performed using as lattice parameters $a=\left(a^{\HH}_{\Exp}/a^{\HH}_{\ThDFT}\right)\times a^{\HH}_{\ThDFT}(P)$
and $c=\left(c^{\HH}_{\Exp}/c^{\HH}_{\ThDFT}\right)\times c^{\HH}_{\ThDFT}(P)$.
Increasing pressure} the anharmonicity  of the lowest energy modes around $M$ and $L$ decreases, but remains relevant even up to 14 GPa. 
A similar conclusion was drawn for 2H-NbSe$_2$, where large anharmonic effects 
and strong temperature dependence of these phonon modes {were} observed as high as 
16 GPa, in a region of its phase diagram
where no CDW transition is observed~\cite{PhysRevB.92.140303}. 


\begin{figure}[t!]
\centering
\includegraphics[width=\columnwidth]{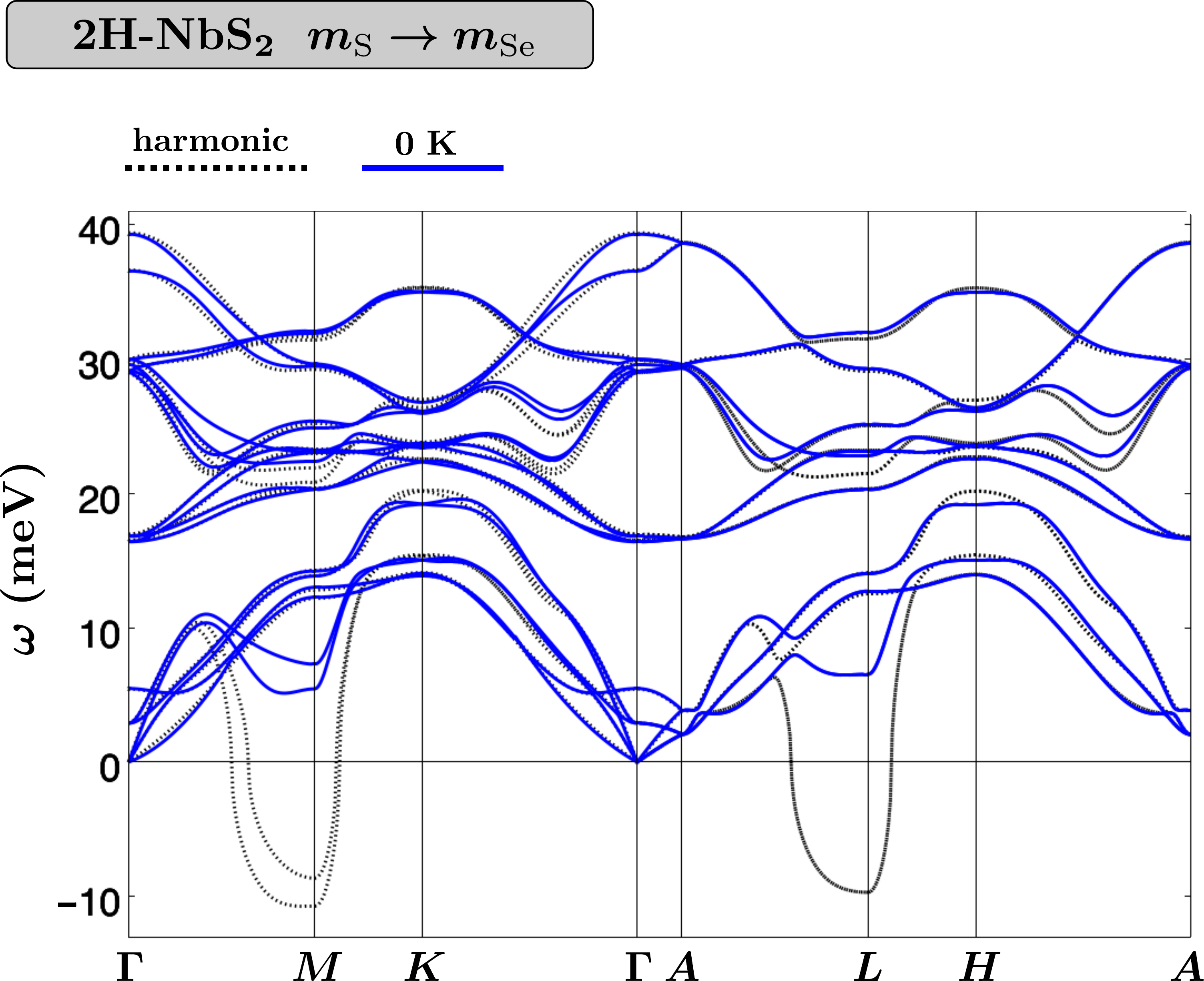}
\caption{(Color online) 2H-NbS$\2$ harmonic phonon dispersion (black
  dashed lines) and SSCHA anharmonic phonon dispersion at $0$~K (blue
  solid lines), at zero pressure, computed replacing the mass of S
  with the mass of Se (more details in the main text). Anharmonicity
  removes the instability also in this case.}
\label{fig:NbS2_massaSe}
\end{figure}

{These results confirm the importance of quantum anharmonicity in 2H-NbS$\2$ 
to describe experimental data and 
the absence of a CDW instability. It is tempting, at this
point, to use the same technique to shed light on the
different CDW behavior exhibited by the very similar
compound 2H-NbSe$\2$. Indeed, as we showed in a previous work \cite{PhysRevB.92.140303},
the SSCHA correctly displays the occurrence of CDW in
2H-NbSe$_2$ at ambient pressure.
One evident difference between 2H-NbS$\2$ and 2H-NbSe$\2$ is, of course, 
the mass of the chalcogen atom. 
We then performed a SSCHA calculation at $0$~K for 2H-NbS$\2$ with ``artificial'' 
S atoms having unaltered electronic configuration but the mass of Se. In other words, we performed a
SSCHA calculation where the average displacements of the atoms from the equilibrium position
is ruled by the Se mass, but for each fixed position of the atoms the 
electronic structure is computed with the normal S atoms. 
The results are shown in Fig.~\ref{fig:NbS2_massaSe}. Also in this case, when quantum anharmonic effects 
are included the system does not show any CDW instability. Thus the different behavior 
of 2H-NbS$\2$ and 2H-NbSe$\2$ cannot be ascribed to a  mass effect but has a more complex origin related to
 the different electron screening on the ions.}


\begin{figure}[t!]
\centering
\includegraphics[width=\columnwidth]{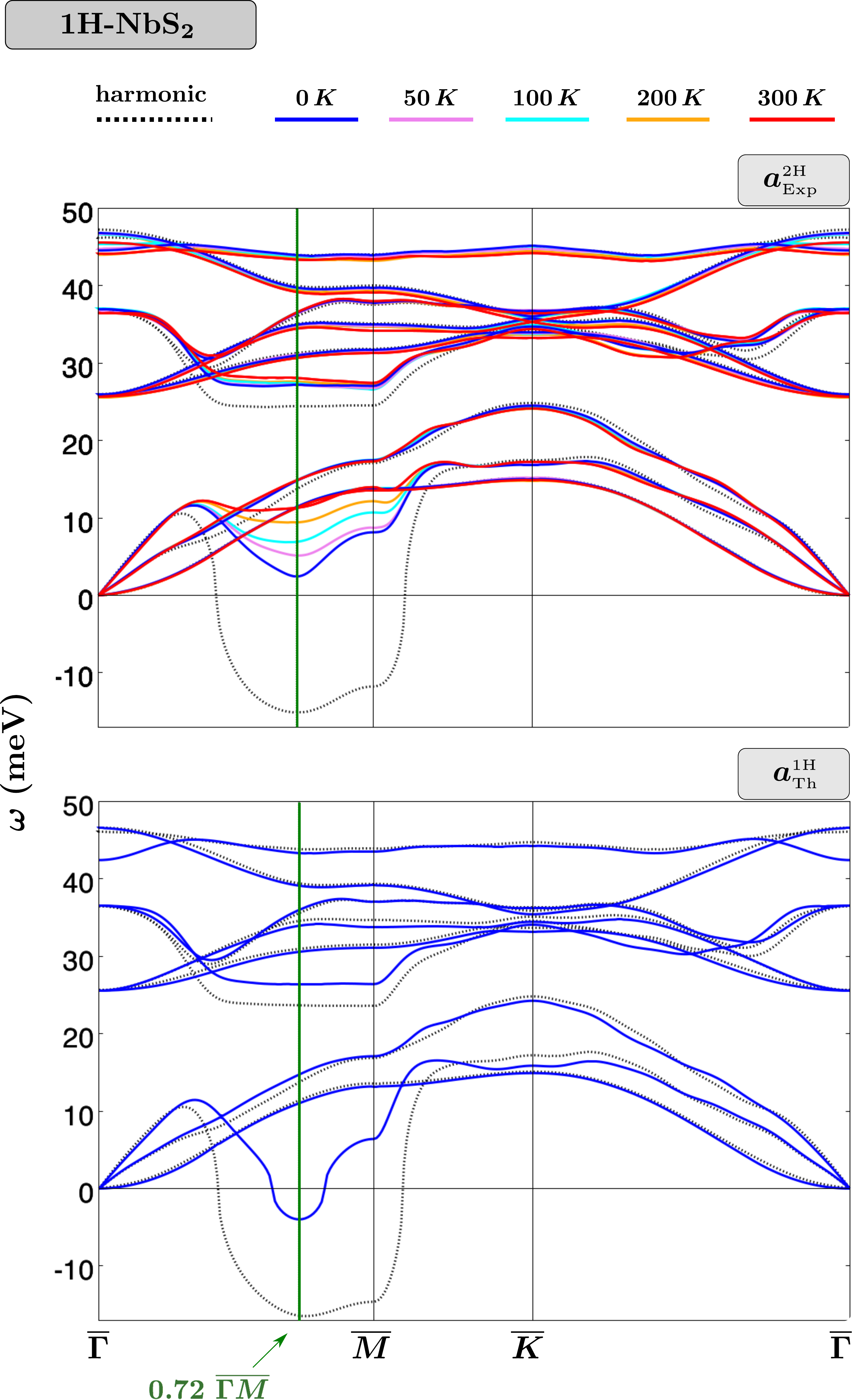}
\caption{(Color online) Suspended 1H-NbS$\2$ harmonic phonon dispersion (black dashed lines) and SSCHA anharmonic phonon dispersion at several temperatures (colored solid lines), at zero pressure. 
Upper panel: results obtained with the experimental in-plane bulk lattice parameter $a^{\HH}_{\Exp}$. The softening of the acoustic mode, localized at $\bq_{\CDW}=0.72\,\overline{\Gamma} \overline{M}$, is more pronounced than in the 2H bulk case.  
However, the frequencies remain real even at $0$~K. Lower panel: results obtained with the theoretical lattice parameter $a^{\mH}_{\Th}$, obtained by fully relaxing the structure taking into account quantum anharmonic effects. 
At $0$~K the frequency at $\bq_{\CDW}=0.72\,\overline{\Gamma} \overline{M}$ becomes imaginary.}
\label{fig:1H-NbS2}
\end{figure}

The validity of the results obtained with the SSCHA method on bulk 2H-NbS$_2$
gives us confidence that a similar calculation on the 1H-NbS$_2$ monolayer may shine light about the effects that dimensionality and environmental conditions (substrate, doping)
can have on the CDW ordering in metallic TMDs.
The suspended 1H-NbS$_2$ monolayer was simulated leaving $12.55\,\angs$ of vacuum space between a 1H layer and its periodic replica.   At conventional static DFT level, we found that the theoretical zero pressure in-plane lattice parameter of the monolayer and the bulk are essentially the same, $a^{\HH}_{\ThDFT}\simeq a^{\mH}_{\ThDFT}\simeq 3.34 \angs$.  Therefore, for the suspended monolayer we use as in-plane lattice parameter the bulk experimental one, $a^{\HH}_{\Exp}=3.33\,\angs$. This value is also compatible with the recent experimental measures $3.29\pm 0.03\,\angs$ and $3.34\,\angs$ reported for the lattice parameter of monolayer grown on substrate in Ref.~\citenum{LinNbS2} and Ref.~\citenum{2019arXiv190103552S}, respectively. 

In the upper panel of Fig.~\ref{fig:1H-NbS2}, we show the harmonic  and SSCHA anharmonic phonon dispersions of  suspended 1H-NbS$\2$ at several temperatures, calculated with 
the lattice parameter $a^{\HH}_{\Exp}$.  As in the bulk case, the system is unstable at harmonic level, but it is stabilized by quantum fluctuations strongly sensitive to the anharmonic potential down to $0$~K. However, comparing Figs. \ref{fig:2H-NbS2_Exp} and~\ref{fig:1H-NbS2}, we observe that even if the used in-plane lattice parameter is the same in both cases, at $0$~K the softest theoretical phonon frequency is approximately  $20\%$ harder in the bulk than in the single layer case, demonstrating that there is a substantial enhancement of the tendency toward CDW in the 2D limit. In the monolayer, the theoretical phonon softening is localized in $\bq_{\CDW}=0.72\,\overline{\Gamma} \overline{M}$, which is quite close to the $\bq_{\CDW}\simeq 2/3\,\overline{\Gamma} \overline{M}$ of the CDW instability experimentally found in 1H-NbS${}_2$ on 6H-SiC(0001)~\cite{LinNbS2} (and in 1H-NbSe$\2$~\cite{Xi2015,Ugeda2015}).  Notice that, since the computed wave-vector of the instability may be affected by the finite grids used in the calculations, we do not discard that it may be slightly  shifted in the infinite grid limit.

Pressure  tends normally to remove CDW ordering. Therefore, considering the proximity of the instability, it cannot be discarded that a tensile dilatation due to the substrate may induce the CDW transition observed for 1H-NbS$\2$ on graphitized 6H-SiC(0001).  However, for the same reason, we cannot exclude the more interesting prospect that the observed CDW be an intrinsic property of this system. Indeed, even small variations of the lattice parameter, compatible with the experimental uncertainly, could have a relevant impact on the results of the calculations, and a more accurate theoretical analysis of the monolayer structure is therefore necessary. As the energy of the soft-mode along $\overline{\Gamma M}$ is of the order of $\simeq 58$~K, 
for a proper analysis of the CDW in the monolayer it is important to fully take into account quantum effects.  Including quantum anharmonic contributions to strain through the technique introduced in Ref.~\citenum{PhysRevB.98.024106}, we find that with the used lattice parameter $a^{\HH}_{\Exp}$  the structure is sligthly compressed, with an in-plane pressure $P=0.66$ GPa.  Upon relaxation we obtain the theoretical lattice parameter  $a^{\mH}_{\Th}=3.35\,\angs$, 
approximatively $0.5\%$ larger than $a^{\HH}_{\Exp}$. 

The harmonic and quantum anharmonic phonons at $0$~K calculated with the lattice parameter $a^{\mH}_{\Th}$ are shown in the bottom panel of Fig.~\ref{fig:1H-NbS2}.  
While at harmonic level the phonon dispersion is not substantially different from the one computed with $a^{\HH}_{\Exp}$, when quantum anharmonic effects are included the
phonon dispersion at $0$~K now shows an instability at $\bq_{\CDW}=0.72\,\overline{\Gamma} \overline{M}$, thus in agreement with the CDW observed for 1H-NbS$\2$ on top of  6H-SiC(0001). The obtained instability is very weak (i.e. the obtained imaginary phonon frequency is very small). Therefore, this result is also compatible with the hypothesis that charge doping from the substrate could be at the origin of the CDW suppression for 1H-NbS$\2$ on top of Au(111), similarly to what it was proposed for the case of 1H-TaS$_2$ on top of Au(111)~\cite{PhysRevB.95.235121}.
Our results show that if quantum anharmonic effects are included, then even a small compression/dilatation of approximately $0.5\%$ removes/induces the charge density wave instability on 1H-NbS$_2$. 
The extreme sensitivity of the CDW on environmental conditions therefore suggests that deposition of 1H-NbS$_2$ on flexible substrates~\cite{Wu, He, Conley}, or a small
charge transfer via field effect, could lead to devices with  dynamical on/off switching of the $3\times 3$ order.

{In conclusion, we have shown that quantum anharmonicity is the key
interaction for the stabilization of the crystal lattice in bulk
2H-NbS$_2$, as it removes the instability found at the harmonic level. 
The calculated temperature dependence of the phonon spectra
are in excellent agreement with inelastic X-ray scattering data.
Anharmonicity remains important even at large pressures.
Given the good agreement between theory and experiment in bulk 2H-NbS$_2$,
we have studied the behavior of the CDW in the 2D limit by
considering single layer 1H-NbS$_2$. We found that suspended 1H-NbS$_2$ undergoes a
quantum phase transition to a CDW state with approximately $3\times 3$
charge ordering in the 2D limit, in agreement with experimental
results on supported samples on  6H-SiC(0001). However, the CDW is extremely
sensitive to environmental conditions, as it is very weak and compressive strains smaller
than $0.5 \%$ are enough to suppress it. This explains the absence of CDW observed
in 1H-NbS$_2$ on top of Au(111). This also suggest that devices with dynamical on/off switching of the $3\times 3$ charge order
can be obtained with deposition of 1H-NbS$_2$ on flexible substrates, or through a small
charge transfer via field effect.
 
\section*{Acnowledgement}
{R.B. acknowledges the CINECA award under the ISCRA initiative (Grant HP10BLTB9A).
Computational
resources were provided by PRACE (Project No. 2017174186) and
EDARI(Grant A0050901202).
I.E. acknowledges financial support from the Spanish Ministry of
Economy and Competitiveness (Grant No. FIS2016-76617-P). 
M.C. acknowledges support from Agence Nationale de la
Recherche under the reference No. ANR-13-IS10-0003-
01. 
We acknowledge support from the Graphene Flagship
(Grant Agreement No. 696656-GrapheneCore1).}

\end{document}